\documentclass[12pt]{article}

\usepackage[numbers, sort&compress]{natbib}
\usepackage{fullpage}
\usepackage{epsfig}
\usepackage{latexsym,amsmath,amssymb,amsthm}
\usepackage{subfig}
\newcommand{\ceil}[1]{\left\lceil {#1} \right\rceil}
\newcommand{\floor}[1]{\left\lfloor {#1} \right\rfloor}

\def\maps11{\stackrel {1-1}{\longmapsto}}

\begin{document}

\title{MAC level Throughput comparison: 802.11ax vs. 802.11ac}

\author{%
Oran Sharon
\thanks{Corresponding author: oran@netanya.ac.il, Tel: 972-4-9831406,
Fax: 972-4-9930525} \\
Department of Computer Science \\
Netanya Academic College \\
1 University St. \\
Netanya, 42365 Israel
\and
Yaron Alpert\\
Intel\\
13 Zarchin St.\\
Ra'anana, 43662, Israel\\
Yaron.alpert@intel.com
\and
Robert Stacey \\
Intel \\
2111 NE 25th Ave, \\
Hillsboro OR 97124 , USA\\
robert.stacey@intel.com\\
}

% \fi %%%%%

\date{}

\maketitle

\begin{abstract} 
In this paper we compare between the maximum Throughputs
received in IEEE 802.11ax and IEEE 802.11ac in a scenario
where a single user continuously transmits to another one. 
The comparison is done as a function of the Modulation/Coding
scheme in use. In IEEE 802.11ax we consider two modes
of operation where in one Acknowledgment frame up to
64 or 256 frames are acknowledged respectively.
IEEE 802.11ax outperforms IEEE 802.11ac by at most 48$\%$ and 29$\%$
in unreliable and reliable channels respectively.
\end{abstract}

\bigskip


\renewcommand{\baselinestretch}{1.3}
\small\normalsize

%%%%%%%%%%%%%%%%%%%%%%%%%%%%%%%%%%%%%%%%%%%%%%%%%%%%%%%%%%%%%%%%

\section{Introduction}

\indent
The latest IEEE 802.11-REVmc Standard (WiFi), created and maintained by 
the IEEE LAN/MAN Standards Committee (IEEE 802.11)~\cite{IEEEBase1}
is currently the most
effective solution within the range of Wireless Local
Area Networks (LAN). Since its first release in 1997,
the standard provides the basis
for Wireless network products
using the WiFi brand, and has since been improved upon
in many ways. One of the main goals of these improvements
is to increase the Throughput achieved by users and to improve
its Quality-of-Service (QoS) capabilities.

To fulfill the promise of increasing
IEEE 802.11 performance and QoS capabilities
a new amendment IEEE 802.11ax, also
known as High Efficiency (HE) was introduced 
recently~\cite{IEEEax}.
IEEE 802.11ax is a six generation type of a WLAN in the IEEE
802.11 set of types of WLANs~\cite{DCC,B}
and it is a successor to IEEE 802.11ac~\cite{IEEEac}.
Currently this project is at a very early stage
of development and 
it is due to be
publicly released in 2019 .
IEEE 802,11ax is predicted 
to have a top capacity of around 10 Gbps and a 
frequency of 2.4 and/or 5 GHz,
and has the goal of providing 4 times
the Throughput of IEEE 802.11ac .
% that
% improves some of the features
% of the PHY and MAC layers of the IEEE 802.11ac in order
% to increase the Throughput of the MAC layer even further. 

In this paper we compare between the Throughputs
of IEEE 802.11ax and IEEE 802.11ac in a scenario
where one user continuously transmits in a single user (SU)
operation mode to another user
without collisions, using aggregation. 
In order to achieve the 4 times Throughput
compared to IEEE 802.11ac, around 10Gbps, the IEEE
802.11ax addresses several new features.

The first feature extends by 4 times the IEEE 802.11ac OFDM
symbols duration while preserving the IEEE 802.11ac Guard Interval (GI)
. In addition, two new Modulation/Coding schemes are introduced
in IEEE 802.11ax, 1024 QAM 3/4 and 1024 QAM 5/6 , MCS10
and MCS11 respectively. In order to support the above
two new features the PHY Preamble in IEEE 802.11ax is longer
than that in IEEE 802.11ac, as we show in Section 2.

Next, in this paper we focus in Two-Level aggregation, first
introduced in IEEE 802.11n~\cite{IEEEBase} and later
extended in IEEE 802.11ac~\cite{IEEEac} 
and IEEE 802.11ax~\cite{IEEEax}. In order to increase
the Throughput in IEEE 802.11ax the MAC acknowledgment window
is extended to 256 MAC Protocol Data Units (MPDU) which
extends the IEEE 802.11ac aggregation capability.

% In this scenario we want
% to evaluate the benefit of
% several new features in the PHY and MAC
% layers of the IEEE 802.11ax that were not investigated yet.
% The first feature concerns the OFDM PHY layer
% where in IEEE 802.11ax there are 4 times bins
% in an OFDM symbol than in IEEE 802.11ac . As a result the
% OFDM symbol in IEEE 802.11ax is longer, However, the same Guard Interval (GI)
% is appended to every symbol in both
% IEEE 802.11ax and 802.11ac resulting
% in a more efficient PHY layer in IEEE 802.11ax .
% On the other hand, the PHY Preamble in IEEE 802.11ax is longer
% than that in IEEE 802.11ac, as we show in Section 2.
% Third, two new Modulation/Coding schemes are
% introduced in IEEE 802.11ax, 1024 QAM 3/4 and 1024 QAM 5/6, MCS10
% and MCS11 respectively.
% Finally, in this paper we focus in Two-Level aggregation, first
% introduced in IEEE 802.11n~\cite{IEEEBase} and later extended in
% IEEE 802.11ac~\cite{IEEEac}. In IEEE 802.11ax it is
% possible to transmit up to 256 MAC Protocol Data Units (MPDU) in
% a single PHY Service Data Unit (PSDU) compared to 64
% in IEEE 802.11ac . Also, in IEEE 802.11ax there is no limit
% on the transmission time of the PHY Protocol
% Data Unit (PPDU - PSDU and its Preamble) while IEEE 802.11ac imposes such a
% limit. In Section 2 we elaborate on the above changes in more
% detail.

In this paper we verify what is the Throughput improvement
achieved in IEEE 802.11ax following the above new features.
In overall, the research on the performance
of IEEE 802.11ax in various scenarios is in
its first steps~\cite{QLYY}.

The paper is organized as follows: In Section 2 we describe in more
details the new features of IEEE 802.11ax mentioned above
and describe the transmission scenario over which
we compare between IEEE 802.11ax and IEEE 802.11ac .
We assume that the reader is familiar with
the basics of the PHY and MAC layers
of IEEE 802.11 described in previous papers, e.g.~\cite{SA}. 
In Section 3 we analytically compute the Throughput
of the transmission scenario described in Section 2 and
in Section 4 we present the Throughputs of the
protocols and compare between them.
In Section 5 we analytically compute the PHY rates 
from which using a 256 MPDUs acknowledgment window size
in IEEE 802.11ax is better than using a 64 MPDUs acknowledgment
window size and finally Section 6 summarizes the paper.
In the rest of the paper we denote IEEE 802.11ax and IEEE 802.11ac
by 11ax and 11ac respectively.

\section{Model}

In this paper we consider the Single User (SU) operation mode
in 11ax vs. that in 11ac.
In this operation mode every transmitted
PHY Protocol Data Unit (PPDU)
is destined to one user only. 
As mentioned, there are several new features
in 11ax compared to 11ac in the PHY and
MAC layers in the SU operation mode.
Assuming an OFDM
based PHY layer,
every OFDM symbol is extended from $3.2 \mu s$ in 
11ac to $12.8 \mu s$ in 11ax. Since
the same Guard Interval (GI) is added to every such symbol,
the overhead in 11ax due to the GI is lower.
Second, in 11ax there are two new 
Modulation/Coding schemes (MCSs), 1024 QAM 3/4 and 1024 QAM 5/6,
MCS 10 and MCS 11 respectively, applicable for bandwidth
larger than 20 MHz. The above
two features enlarge the PHY rate of 11ax .

In this paper we focus in the Two-Level aggregation
scheme, first introduced
in IEEE 802.11n~\cite{IEEEBase}, in which several 
MPDUs are transmitted in a 
single PHY Service Data Unit (PSDU). 
Such a PSDU
is denoted Aggregate MAC Protocol Data Unit (A-MPDU) frame.
In Two-Level aggregation 
every MPDU contains several MAC Service Data Units (MSDU). 
MPDUs are separated 
by an MPDU Delimiter field of 4 bytes and each MPDU contains
MAC Header and Frame Control Sequence (FCS) fields.
MSDUs within an MPDU
are separated by a SubHedaer field of 14 bytes. Every MSDU
is rounded to an integral multiply of 4 bytes
together with the SubHeader field. Every MPDU is also
rounded to an integral multiply of 4 bytes.
In 11ax and 11ac the size of an MPDU
is limited to 11454 bytes. In 11ac an A-MPDU
is limited to 1048575 bytes and this limit is
removed in 11ax . In both 11ac and
11ax the transmission time of the PPDU (PSDU and
its Preamble) is limited to $\sim 5.4ms$ ($5400 \mu s$) 
due to L-SIG (one of the legacy
Preamble's fields) duration limit~\cite{IEEEBase1}.
.

% In the MAC layer IEEE 802.11ax enables to transmit
% MPDUs from several Traffic Streams (TS) in a single A-MPDU
% frame, a feature that
% does not exist in IEEE 802.11ac where MPDUs of only a single
% TS are allowed.
In this paper we also assume
that all the MPDUs transmitted in an A-MPDU
frame are from the same Traffic Stream (TS).
In this case up
to 256 MPDUs are allowed in an A-MPDU frame of 11ax,
while
in 11ac up to only
64 MPDUs are allowed.

\begin{figure}
\vskip 5cm
\includegraphics{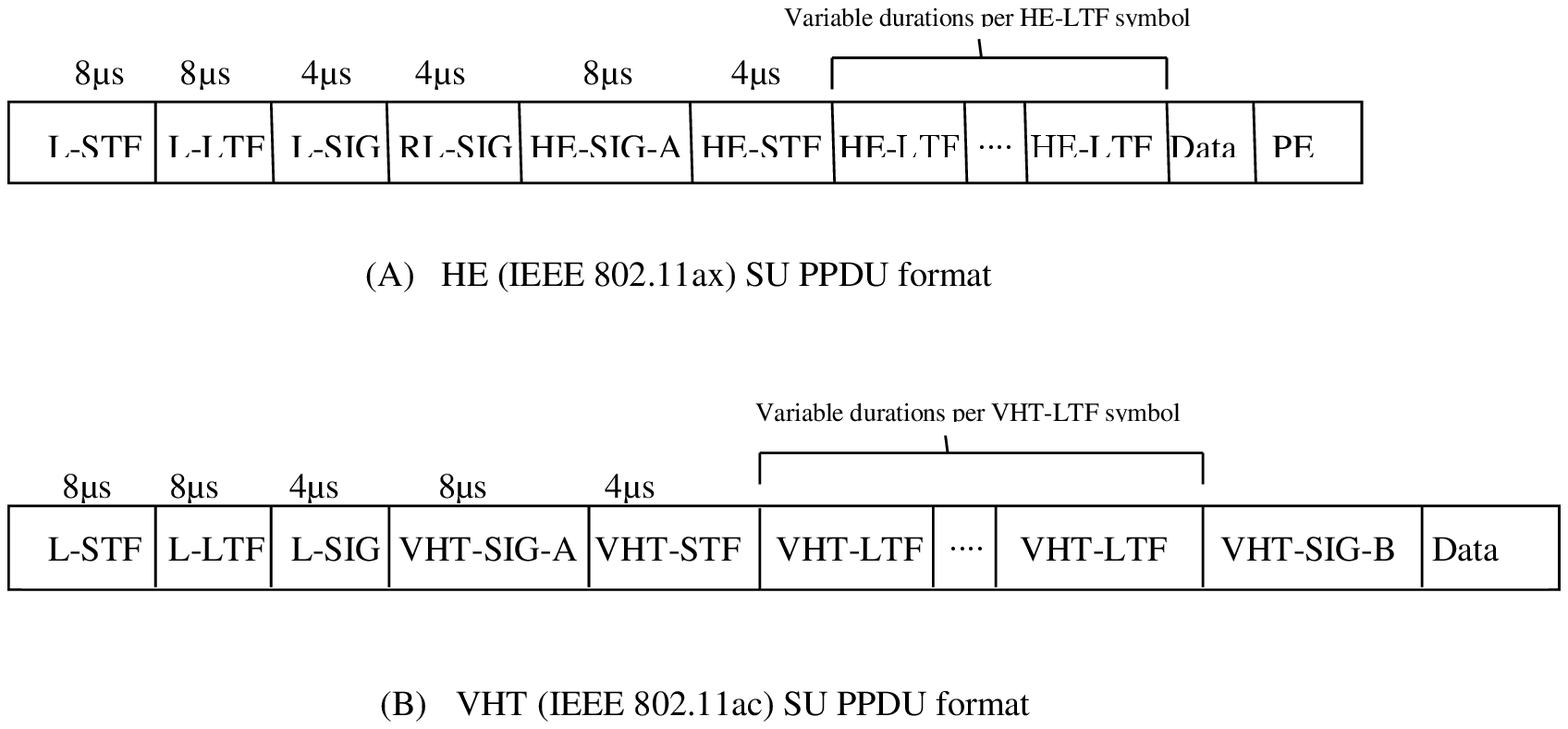}
\caption{The PPDU format in Single User (SU) mode in VHT and HE.}
\label{fig:PPDUformat}
\end{figure}

In Figure~\ref{fig:PPDUformat} we show the PPDU formats
in 11ax and 11ac in parts (A) and (B) respectively.
In the 11ax PPDU format there are HE-LTF fields,
the number of which equals to the number of Spatial Streams (SSs)
in use. In this paper we assume that each such field is of the shortest
length possible, i.e. 
$7.2 \mu s$~\cite{IEEEax}.
In the PPDU format of 11ac there are
the VHT-LTF fields, the number of which equals
again to the number of SSs, and each is $4 \mu s$.
Notice that in SU mode and when using the
same number $S$ of SS, the
Preamble in 11ax is longer than that
in 11ac by $S \cdot (7.2-4)=S \cdot 3.2 \mu s$.

Notice also that the PSDU frame in 11ax contains
a Packet Extension (PE) field.
This field is mainly used in Multi-User (MU) mode
and so we assume that it does not present, i.e. 
it is of length $0 \mu s$.

\begin{figure}
\vskip 4cm
\includegraphics{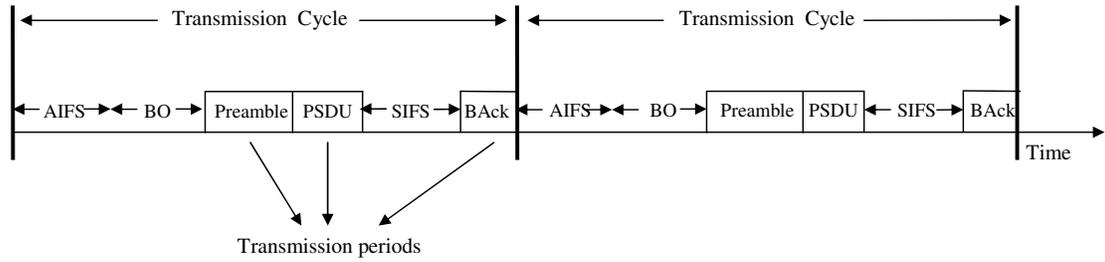}
\caption{The UDP like traffic pattern.}
\label{fig:UDPtraffic}
\end{figure}

We also assume a UDP like traffic where the AP continuously
transmits Data MSDUs to a station, and the station responds with
the BAck control frame.
A transmission of a PPDU from the AP followed by a BAck
control frame from the station is denoted {\it Transmission Cycle}
and such a cycle repeats itself continuously, 
as shown in Figure~\ref{fig:UDPtraffic}.
We also assume the compressed BAck
frame format and consider two cases: in one case the AP transmits
up to 64 MPDUs in every A-MPDU frame and so the BAck
frame is 32 bytes long. It contains 8 bytes, i.e. 64 bits,
each acknowledging one MPDU. In the second case, that is
relevant to 11ax only,
the AP can transmit up to 256 MPDUs in an A-MPDU
frame and so the BAck frame is 56 bytes long, 
containing 32 bytes for acknowledging MPDUs.
The BAck frame is transmitted in legacy mode
using a 24 Mbps PHY rate. Therefore, its transmission
times are $31 \mu s$ and $39 \mu s$ in the above two
cases respectively. 

Finally, we consider several channel conditions which
are expressed by different values of the Bit Error Rate (BER)
which is the probability that a bit arrives successfully
at the destination. We assume a model where these probabilities
are independent from bit to bit~\cite{L1}.

\section{Throughput computation}

Let $X$ be the number
of MPDU frames in an A-MPDU frame, numbered $1,..,X$, and $Y_i$ be the number
of MSDUs in MPDU number $i$.

Also, let 
$O_P = AIFS+BO+Preamble+SIFS+BAck$, 
$O_M = MPDU Delimiter+MacHeader+FCS$,
$Len=4 \cdot \ceil{\frac{L_{DATA}+14}{4}}$ 
and 
$C_i = 8 \cdot 4 \cdot \ceil{\frac{O_M+Y_i \cdot Len}{4}}$.

Then, the Throughput in both 11ax and 11ac is given by
Eq.~\ref{equ:thrtwole}~\cite{SA}:

\begin{equation}
Thr=
\frac
{8 \sum_{i=1}^{X} \cdot Y_i \cdot L_{DATA} \cdot (1-BER)^{C_i}}
{O_P + TSym \ceil{\frac{\sum_{i=1}^{X} \cdot C_i + 22}{TSym \cdot R }}}
\label{equ:thrtwole}
\end{equation}

\normalsize

$TSym$ is the length of an OFDM symbol and every transmission
must be of an integral number of OFDM symbols.
The additional 22 bits in the denomination
are due to the SERVICE and TAIL fields that are added to every
transmission by the PHY layer conv. protocol~\cite{IEEEBase1}.

The function in Eq.~\ref{equ:thrtwole} is not continuous and so
it is difficult to find the optimal X and Y. However, in~\cite{SA} it is
shown that if one neglects the rounding in the
denomination of Eq.~\ref{equ:thrtwole} then the optimal
solution has the property that all the MPDUs
contain almost the same number of MSDUs: the difference
between the largest and smallest number of MSDUs in MPDUs
is at most 1. The difference is indeed 1
if the limit on the transmission time of the PPDU
does not enable to transmit
the same number of MSDUs in all the MPDUs.

If one neglects the rounding of the denomination 
of Eq.~\ref{equ:thrtwole} the received Throughput
for every X and Y is as large as that received
in Eq.~\ref{equ:thrtwole}. The difference depends
on the size of the denomination.

We therefore use the result in~\cite{SA} and look for the
maximum Throughput as follows: We check for every
X, $1 \le X \le 64$ (also $1 \le X \le 256$ for 11ax)
and for every Y, $1 \le Y \le Y_{max}$, what is the
received Throughput such that $Y_{max}$ is the
maximum possible number of MSDUs in an MPDU.
All is computed taking into account
the upper limit of $5.4 ms$ on the transmission time
of the PPDU (PSDU+Preamble). In case where it is not possible
to transmit the same number of MSDUs in all
the MPDUs, part of the MPDUs have one more MSDU
than the others, up to the above upper limit
on the transmission time. We found that the smallest
denomination of any of the maximum Throughputs
is around $1000 \mu s$. Neglecting the rounding
in the denomination
reduces its size by at most $13.6 \mu s$ in 11ax
and $4 \mu s$ in 11ac. Thus, the mistake in the
received maximum Throughputs is in the order of
at most 1.4$\%$.

\section{Throughput comparison between IEEE 802.11ax and IEEE 802.11ac}

\begin{figure} 
\vskip 3cm
\includegraphics{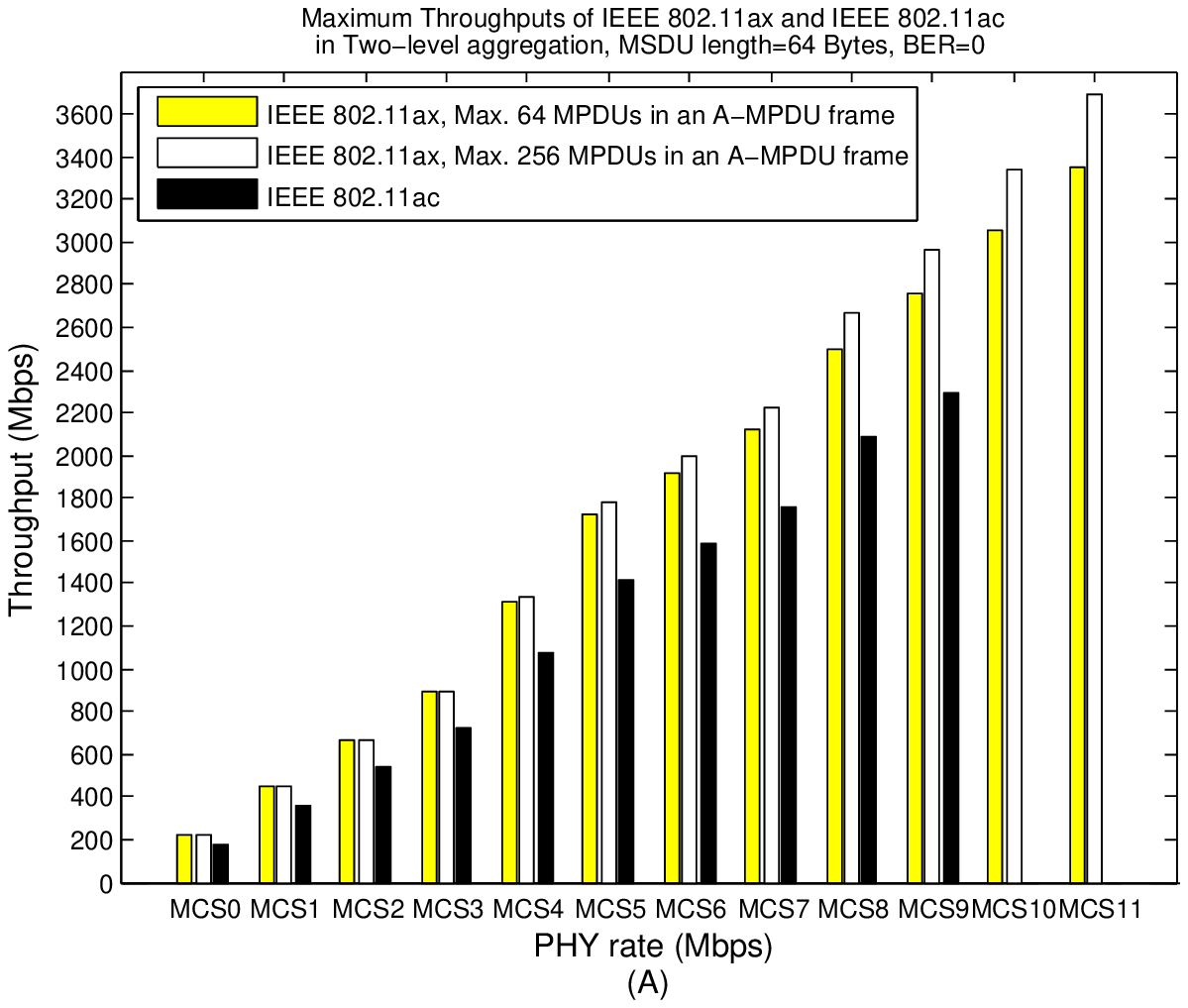}
\includegraphics{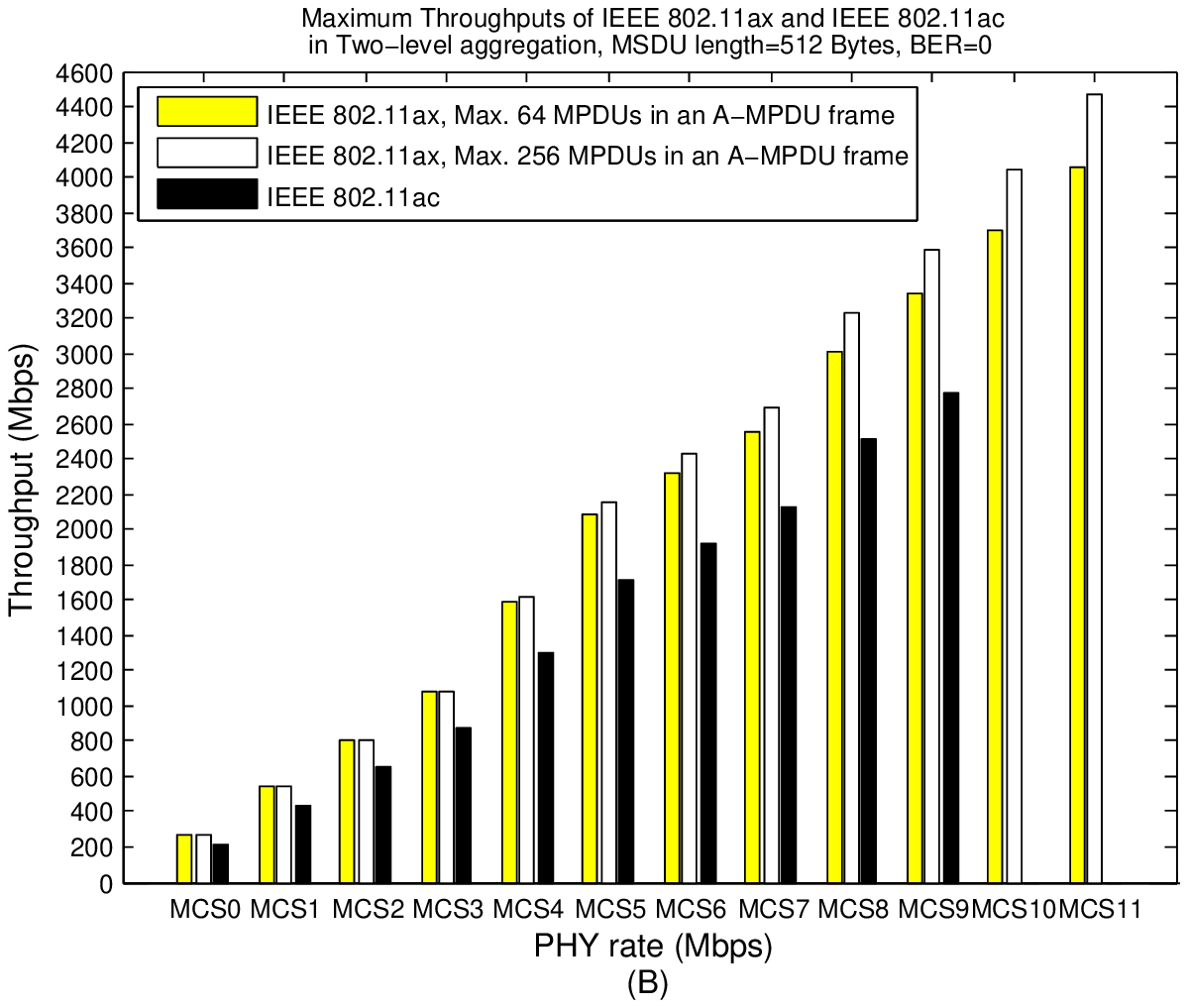}
\includegraphics{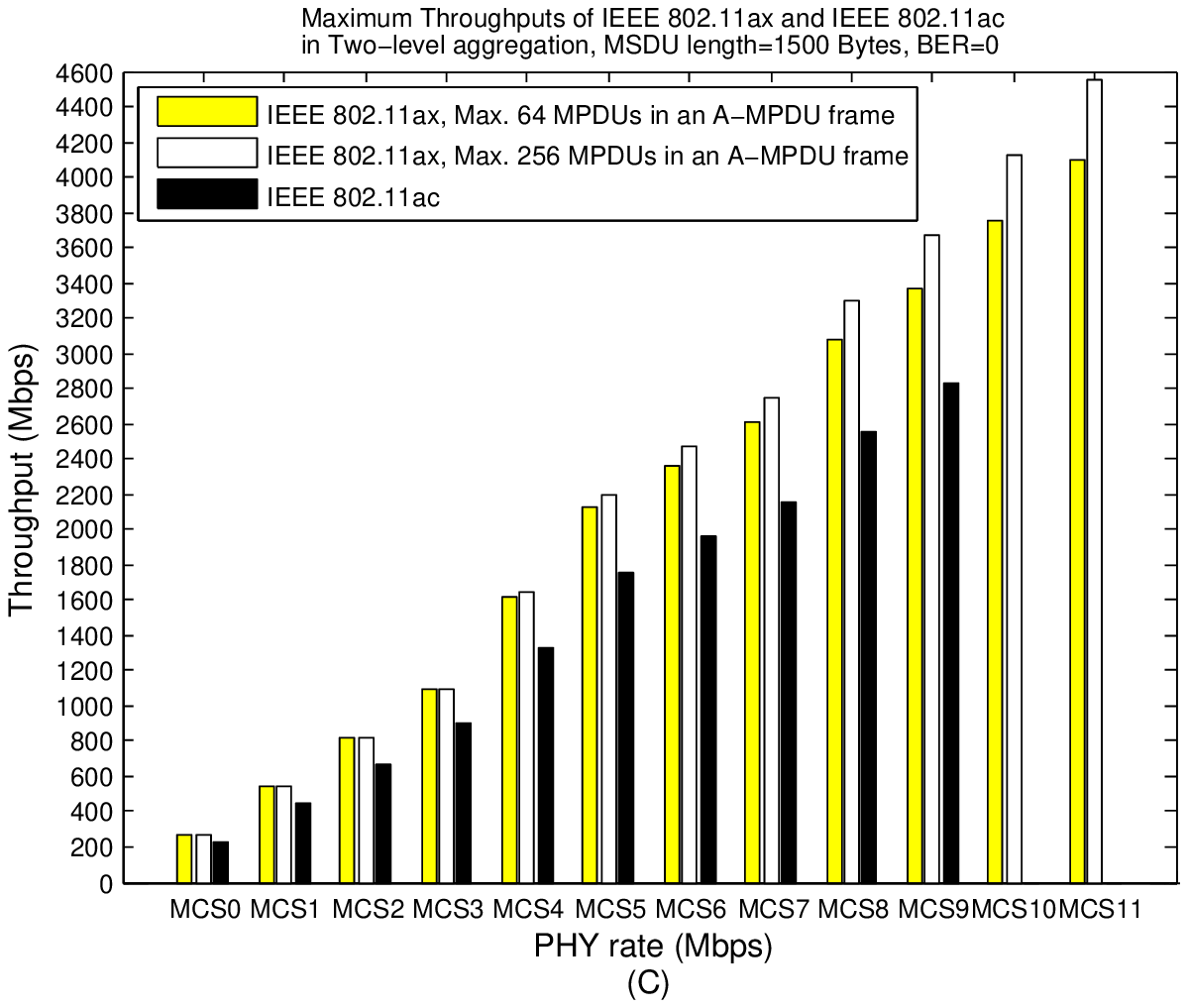}
\caption{Comparison between the maximum Throughputs of 802.11ax and 802.11ac in
the Two-level aggregation scheme, single user operatopm mode and different length MSDUs. BER=0.}
\label{fig:comptwole}
\end{figure}

\begin{figure}
\vskip 5cm
\includegraphics{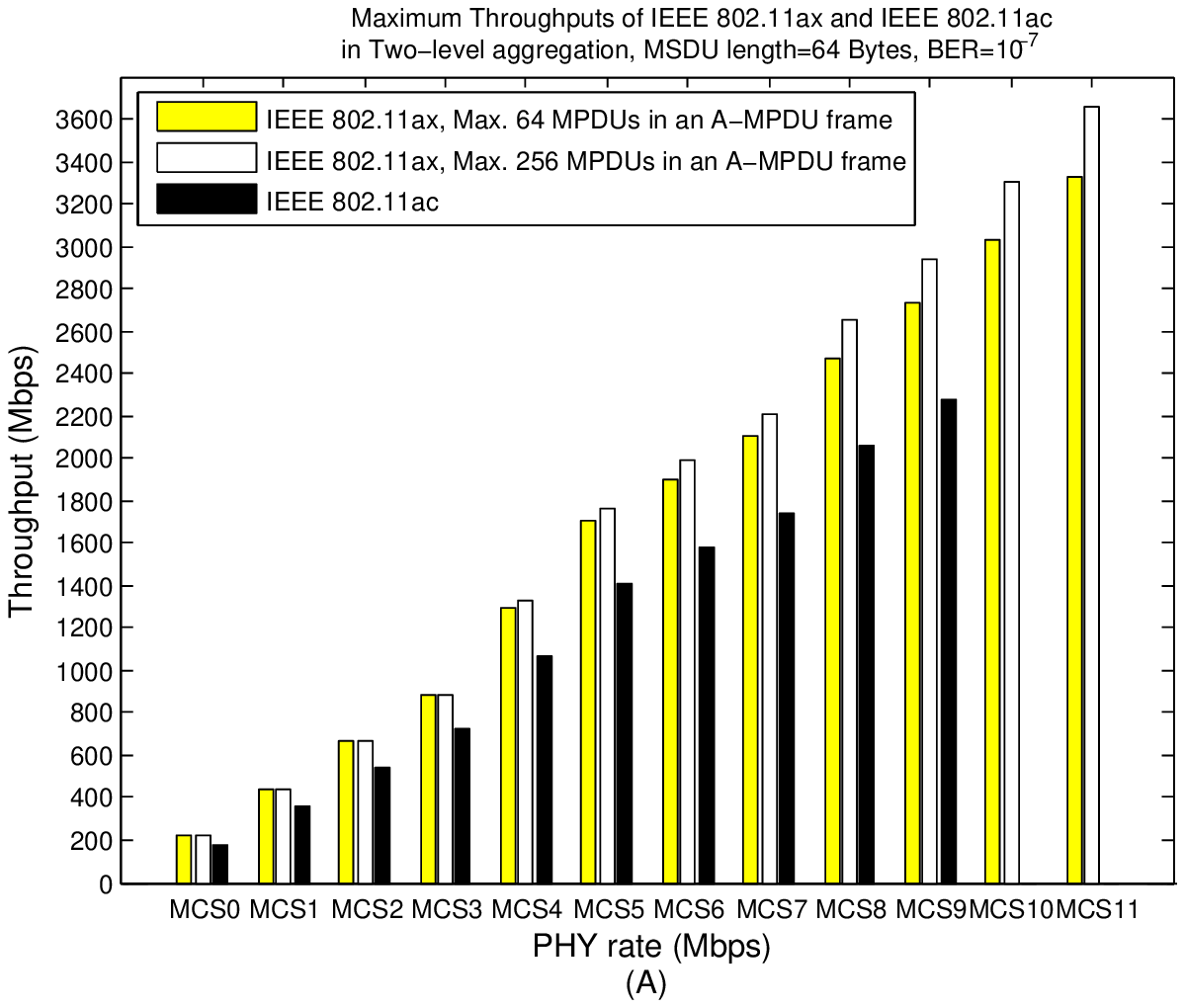}
\includegraphics{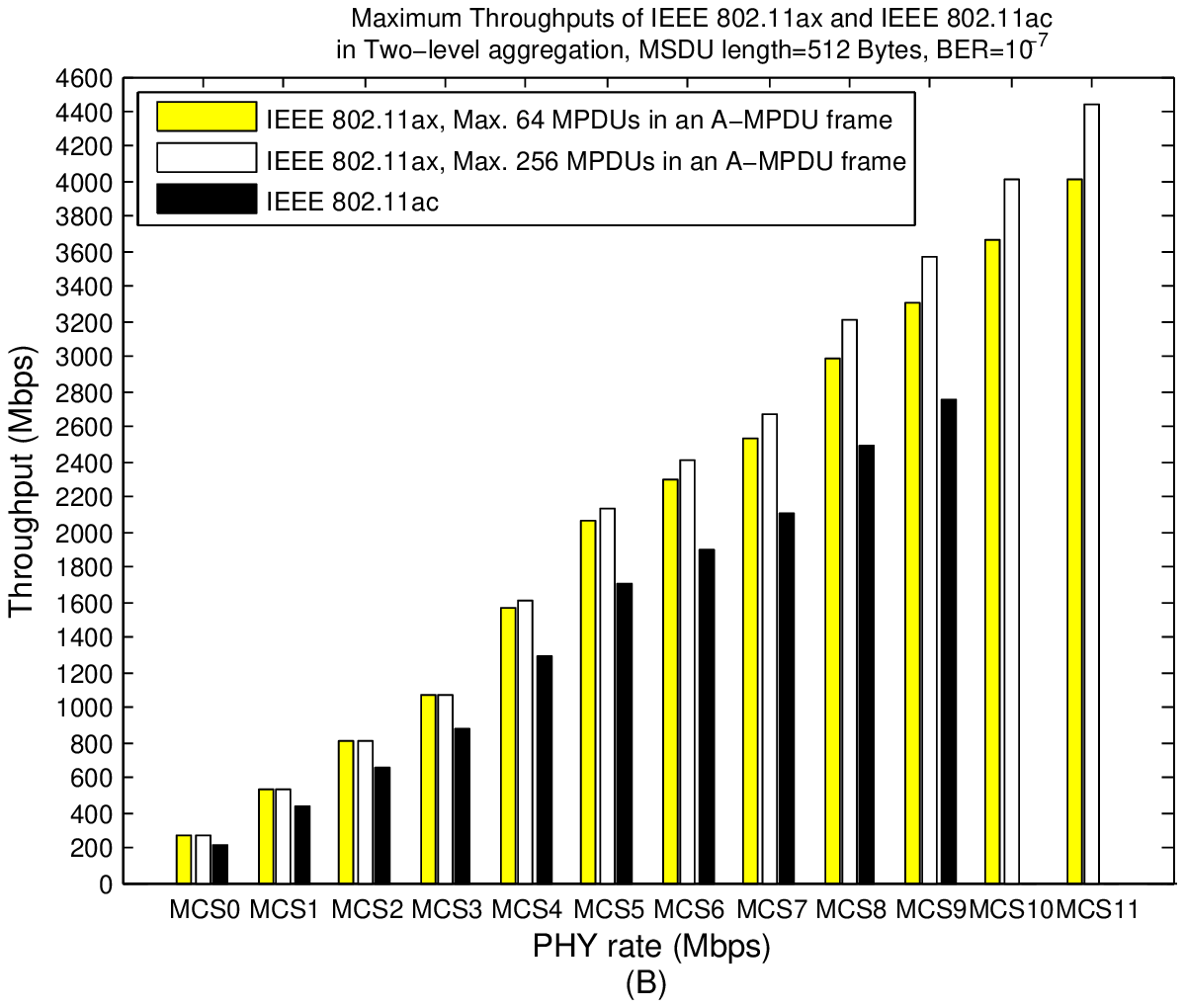}
\includegraphics{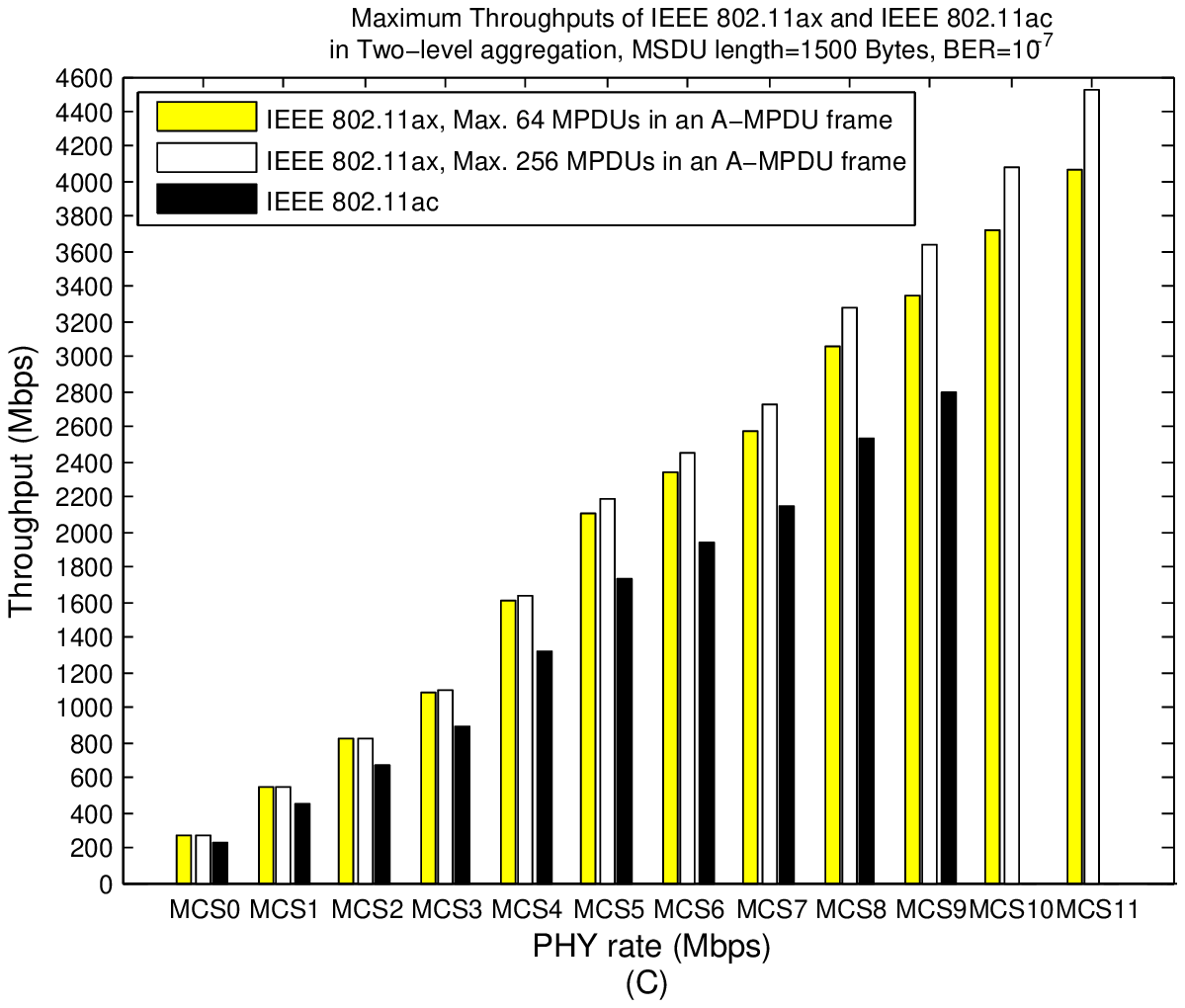}
\caption{Comparison between the maximum Throughputs of 802.11ax and 802.11ac in
the Two-level aggregation scheme, single user operatopm mode and different length MSDUs. BER=$10^{-7}$.}
\label{fig:comptwole5}
\end{figure}

\begin{figure}
\vskip 5cm
\includegraphics{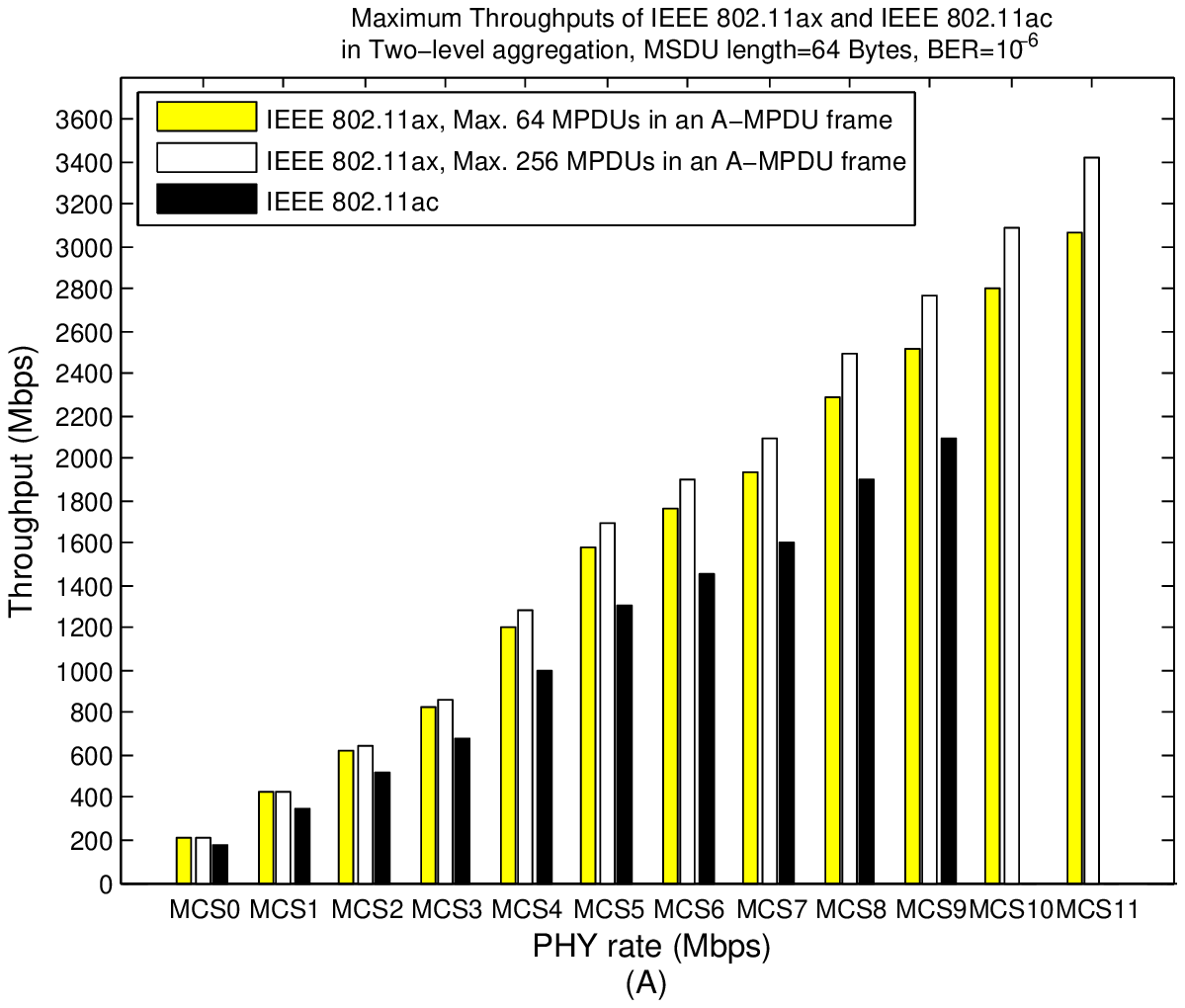}
\includegraphics{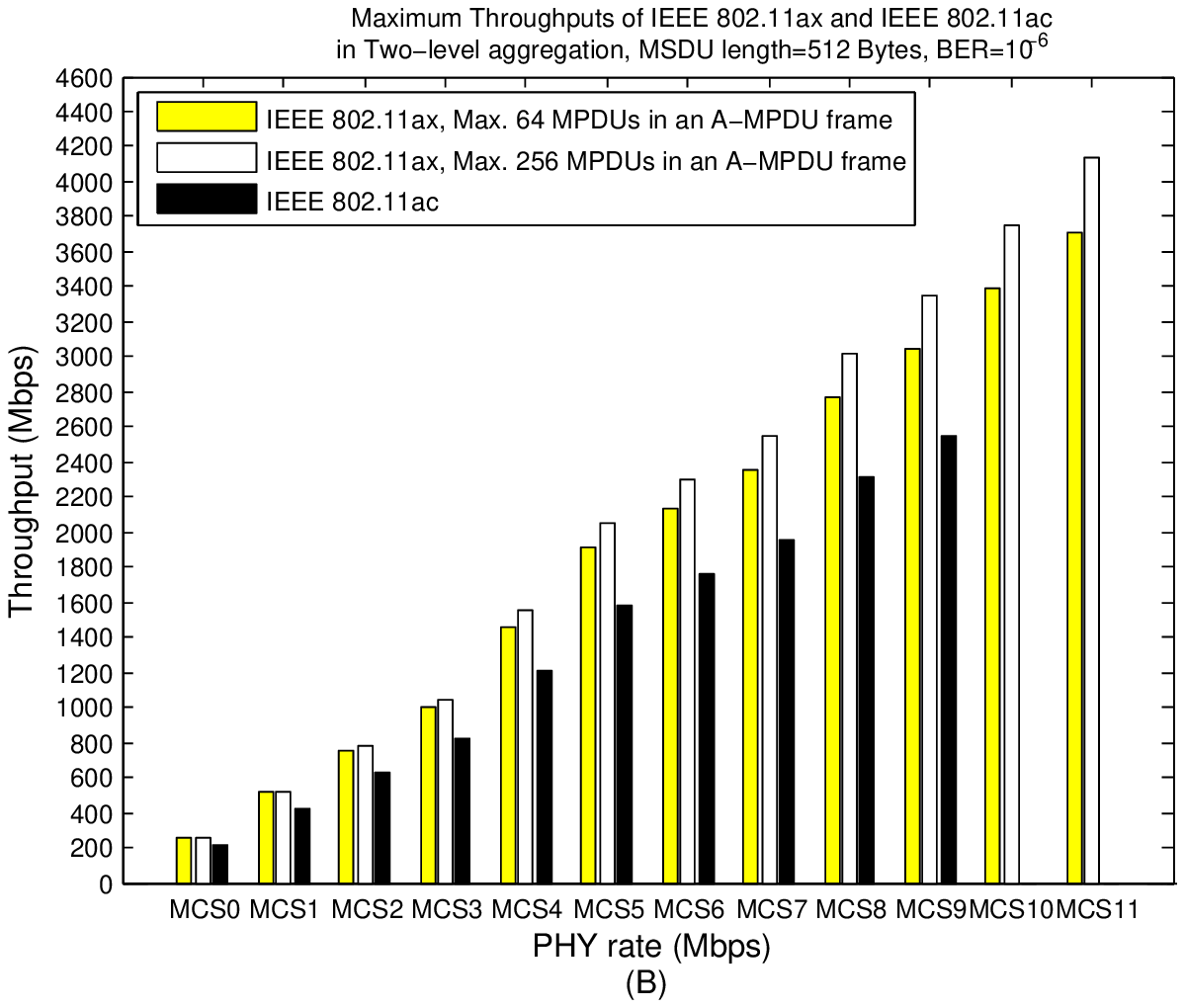}
\includegraphics{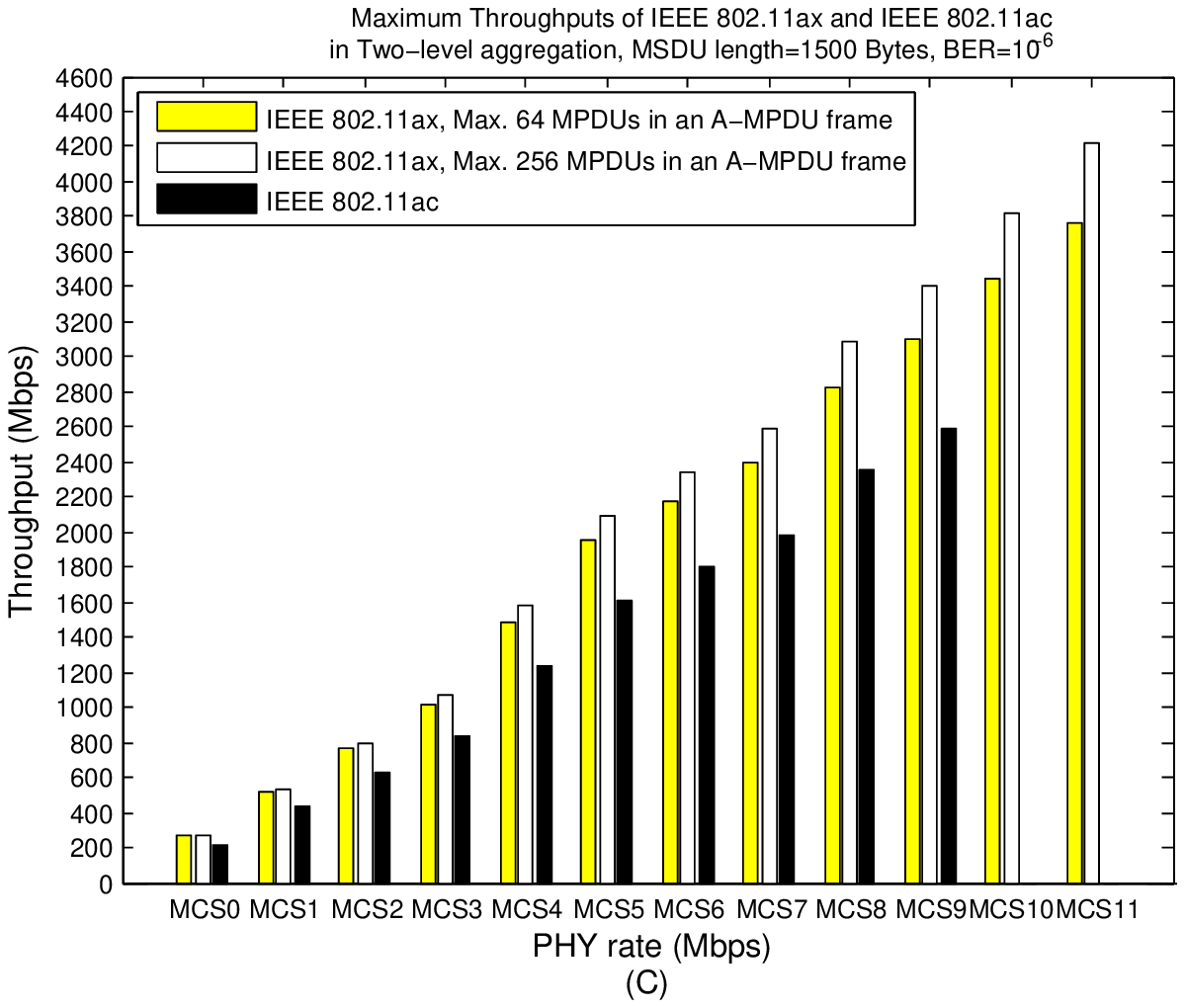}
\caption{Comparison between the maximum Throughputs of 802.11ax and 802.11ac in
the Two-level aggregation scheme, single user operatopm mode and different length MSDUs. BER=$10^{-6}$.}
\label{fig:comptwole6}
\end{figure}

\begin{figure}
\vskip 5cm
\includegraphics{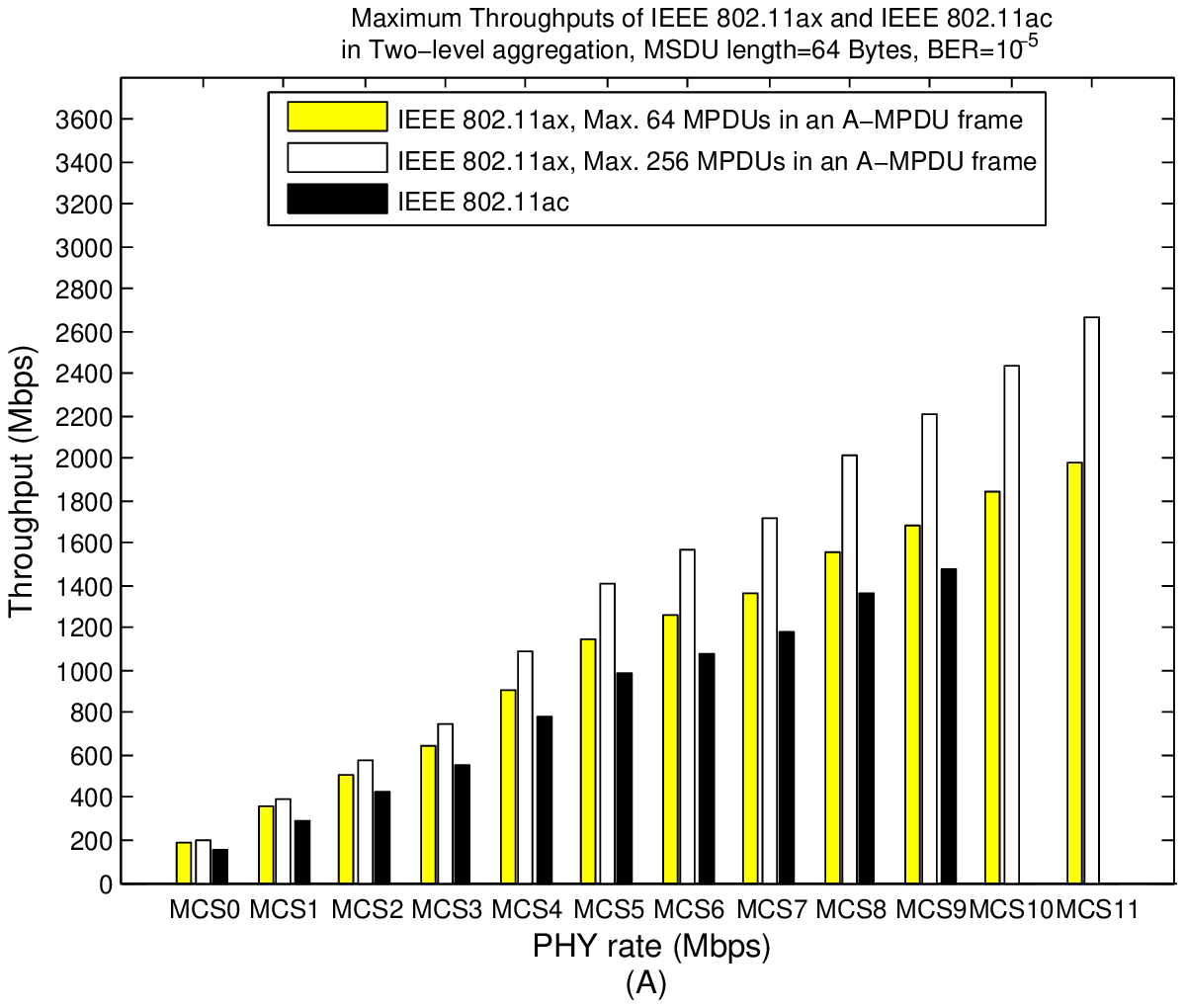}
\includegraphics{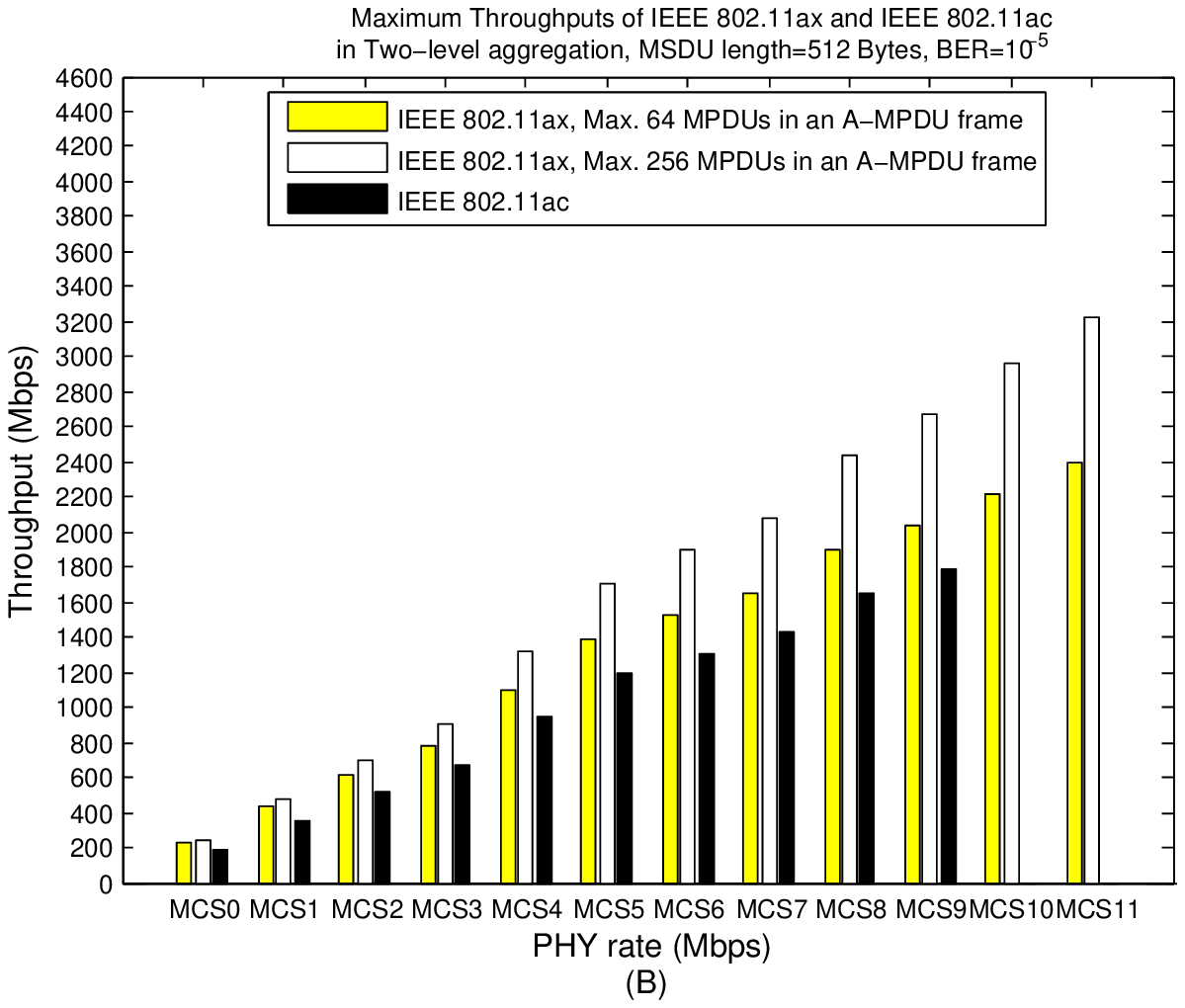}
\includegraphics{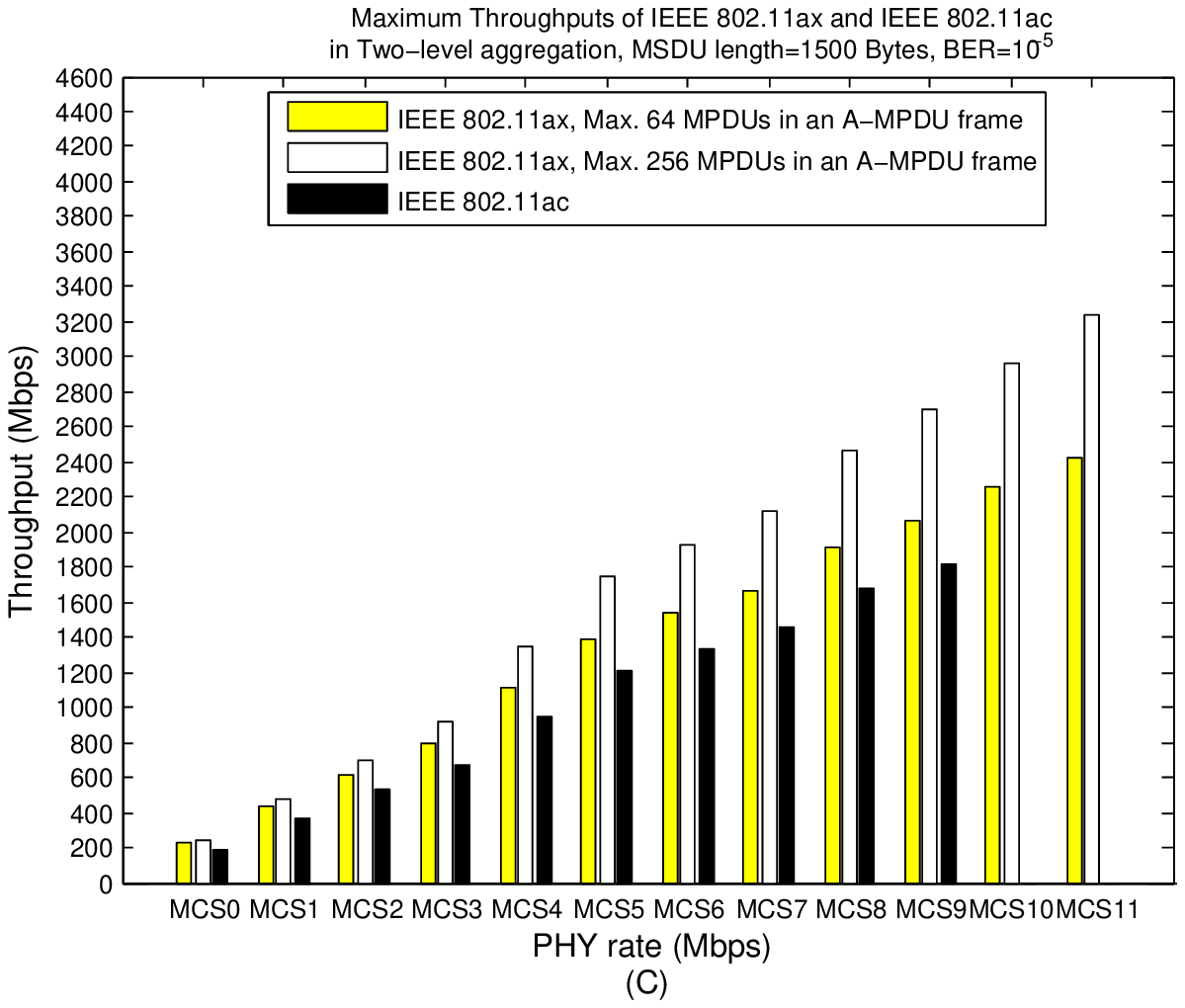}
\caption{Comparison between the maximum Throughputs of 802.11ax and 802.11ac in
the Two-level aggregation scheme, single user operatopm mode and different length MSDUs. BER=$10^{-5}$.}
\label{fig:comptwole7}
\end{figure}

In Figures~\ref{fig:comptwole},~\ref{fig:comptwole5},~\ref{fig:comptwole6},~\ref{fig:comptwole7} 
we show the maximum Throughputs of 11ax
and 11ac for four different channel's
conditions: BER$=0, 10^{-7}, 10^{-6}, 10^{-5}$
respectively. Every figure contains results
for 3 different
sizes $L_{DATA}$ of MSDUs: $L_{DATA}=64, 512$ and
$1500$ octets in parts (A), (B) and (C)
respectively. There are results for 11ac,
with  64 MPDUs in every A-MPDU frame,
for 11ax with  64 MPDUs in every A-MPDU
frame and for 11ax with 256 MPDUs in every
A-MPDU frame. The last two flavors of 11ax
are denoted 11ax/64
and 11ax/256 respectively.

First notice that in every figure
the Throughput is shown as a function of
the MCSs in the x-axis. In every MCS
11ax and 11ac enable different PHY rates
and so the comparison criteria is the Throughput
of the two protocols 
in every MCS in use. Also notice that MCS 10 and MCS 11 are
not possible in 11ac and so 
11ac does not have results for these MCSs.
In 11ac the PHY rates for MCS0-MCS9 are
234, 468, 702, 936, 1404, 1872, 2106, 2340, 2808 and 3120 Mbps
respectively, assuming a 160MHz channel,
4 SSs and a $0.8 \mu s$ Guard Interval.
In 11ax the PHY rates for MCS0-MCS11 are
288, 576, 864, 1152, 1729, 2305, 2594, 2882, 3458, 3843, 4323
and 4803 Mbps respectively.

In all the figures the  performance of 11ax is better
than that of 11ac. This is due to the larger PHY rates
that 11ax enables in every MCS compared to 11ac. 
For BER$=$0 11ax/256 outperforms 11ac by 29$\%$
and in BER$=10^{-5}$ the improvement reaches 48$\%$.
When comparing between 11ax/64 and 11ax/256 one
can see that for BER$=$0 11ax/256 outperforms 11ax/64 only for
MCSs higher than MCS2. On the other hand 
in the case of BER$=10^{-5}$ 11ax/256 outperforms 11ax/64
starting from MCS0. The reason for this difference
is as follows: for BER$=$0 it is worth to transmit
MPDUs with as much MSDUs as possible. Thus,
not many MPDUs are transmitted when the maximum Throughput
is received and the limiting parameter on the Throughput
is the limit on the PPDU transmission time. Therefore,
is small PHY rates, i.e. small MCSs, 11ax/256 has no advantage
over 11ax/64. Only when the PHY rates increase, the limit
of 64 MPDUs in 11ax/64 begins to be significant and 11ax/256
begins to outperform 11ax/64. When BER$=10^{-5}$ it is worth
to transmit short MPDUs because the failure
probability of an MPDU increases with its length.
In small PHY rates the limiting parameter is now
the number of MPDUs and not the limit on the
PPDUs' transmission times. Therefore, 11ax/256 outperforms
11ax/64 also in small indexed MCSs.

Notice that 11ax/256 outperforms 11ac in BER$=10^{-5}$, in percentage,
more than in BER$=$0. The main overhead
incurred in the transmissions is $O_P$. In BER$=$0
MPDUs are large with relatively many MSDUs. On the
other hand in BER$=10^{-5}$ MPDUs are short in order to keep
on large transmission success probabilities.
More MPDUs in BER$=10^{-5}$ are therefore more
significant than in BER$=$0 and so is the
relative improvement in Throughput between 11ax/256 and 11ac .

\section{Acknowledgment window size analysis}

One can conclude from the results in Section 4 two findings:
First, as the BER is smaller, 11ax/256 outperforms
11ax/64 from larger PHY rates. Second, the MCS from which
11ax/256 outperforms 11ax/64 is not dependent on the MSDU size.
We want to investigate these phenomena further.

In the following analysis we use the above mentioned
approximation from~\cite{SA}
where we neglect the rounding in the
denomination of Eq.~\ref{equ:thrtwole} and assume
that all the MPDUs contain the same number of MSDUs.
We also neglect the rounding of the MPDU size
and the addition of the 22 bits in the denomination.
Following this approximation Eq.~\ref{equ:thrtwole} 
turns out to be Eq.~\ref{equ:tlthrerr}:

\begin{equation}
Thr=
\frac
{8 \cdot  X \cdot Y \cdot L_{DATA} \cdot (1-BER)^{8 \cdot (O_M+Y \cdot Len)}}
{O_P +  \frac{8 \cdot X \cdot (O_M+Y \cdot Len)}{R}}
\label{equ:tlthrerr}
\end{equation}

Notice from Eq.~\ref{equ:tlthrerr} that
given a number $Y$ of MSDUs in an MPDU, it is
worthwhile to contain as many MPDUs as possible
in the A-MPDU frame, up to the limit on the PPDU transmission
time.

\subsubsection{Reliable channel, BER$=$0}

Let MCS$_C$ be the MCS from which 11ax/256 outperforms
11ax/64.  For BER$=$0 it is possible to compute MCS$_C$
accurately. Recall that $O_M$ is the sum of the lenghs of
the MAC Header, MPDU Delimiter and FCS fields in bytes.
Also recall that $Len=4 \cdot \ceil{\frac{L+14}{4}}$, 
let $P_r$ be the length
of the Preamble in $\mu s$ ($64.8 \mu s$ in our case),
$R$ be the PHY rate and $T$ be the
limit on the transmission time of the PPDU ( 5400 $\mu s$ in
our case ). Finally, let $Y_{max} = \floor{\frac{11454-O_M}{Len}}$
be the maximum possible number of MSDUs per MPDU frame.
For BER$=$0 it is most efficient to include $Y_{max}$ MSDUs
per MPDU frame and as many MPDUs in the A-MPDU frame
up to the limit $T$.
Then, one receives the following equation for 11ax/64 assuming
that the PHY rate enables to transmit 64 MPDUs of $Y_{max}$ MSDUs
each: $T = \frac{64 \cdot (O_M +Y_{max} \cdot Len)}{R}+P_r$.
The largest PHY rate the enables the transmissions of up to 64 MPDUs
is $R=\frac{64 \cdot (O_M+Y_{max} \cdot Len)}{T-P_r}$.

For $L_{DATA}=1500$ bytes ($Len=1516$ bytes) it turns out
that $R=1021 Mbps$. Neglecting the rounding of $Y_{max}$
one receives that $R=\frac{64 \cdot 11545 \cdot 8}{T-P_r}$ which, 
independently of $L_{DATA}$, equals 1099 Mbps for $T=5400 \mu s$ and
$P_r = 64.8 \mu s$.  The range 1021-1099 Mbps falls between
MCS2 and MCS3 i.e. 11ax/256 outperforms 11ax/64 starting
from $MCS_C = MCS3$ for any
MSDU length $L_{DATA}$ up to 1500 bytes.
In Figure~\ref{fig:comptwole} the difference between
11ax/64 to 11ax/256 in MCS3 is too small to be
noticed, however from MCS4 the difference is noticeable.

\subsubsection{Unreliable channel, BER$>$0}

For positive BERs the optimal number of MSDUs
per MPDU is not necessarily $Y_{max}$. 
Therefore, we use the following approximation.
Given that it is worthwhile to transmit as long PPDUs as possible,
then let $X_{opt}$ and $Y_{opt}$
be the number of MPDUs and the number of
MSDUs per MPDU
respectively in the optimal A-MPDU, i.e. the A-MPDU that achieves the
largest Throughput. Then, 
Eqs.~\ref{equ:tl1} and~\ref{equ:tl2} can give a relation between
$X_{opt}$ and $Y_{opt}$:

\begin{equation}
T=
\frac
{X_{opt} \cdot ( Y_{opt} \cdot Len + O_M )}
{R}
+ Pr
\label{equ:tl1}
\end{equation}

Or:

\begin{equation}
Y_{opt}=
\frac
{R \cdot T - R \cdot Pr - X_{opt} \cdot O_M}
{X_{opt} \cdot Len}
\label{equ:tl2}
\end{equation}

Using 
Eqs.~\ref{equ:tl1} and~\ref{equ:tl2}
the search for the optimal A-MPDU
can consider only the number $X$ of MPDUs and the
number $Y$ of MSDUs per MPDU that maintain
Eq.~\ref{equ:tl2}. Eq.~\ref{equ:tlthrerr}
can therefore be re-written as: 

\begin{equation}
Thr=
\frac
{8 \cdot  X \cdot  
(
\frac
{R \cdot T - R \cdot Pr - X \cdot O_M}
{X \cdot Len}
)
\cdot L_{DATA} \cdot (1-BER)^{8 \cdot (O_M+
(
\frac
{R \cdot T - R \cdot Pr - X \cdot O_M}
{X \cdot Len}
)
\cdot Len)}}
{O_P -Pr + T }
\label{equ:tl3}
\end{equation}

Notice that the denomination of Eq.~\ref{equ:tl3}
is constant because we use the outcome that it is
most efficient that the transmission time of the
PPDU will be the largest possible.

To find the largest Throughput we derive Eq.~\ref{equ:tl3}
according to $X$ and find that the optimal X is the single 
positive solution 
of a quadratic equation, which reveals that Eq.~\ref{equ:tl3}
is unimodal.
The optimal $X$, $X_{opt}$, is given by Eq.~\ref{equ:tl4}:

\begin{equation}
X_{opt}=
\frac
{R \cdot (T - Pr) \cdot ln(1-BER) \cdot O_M}
{2}
\cdot
(1-\sqrt{1-\frac{4}{O_M \cdot ln(1-BER)}})
\label{equ:tl4}
\end{equation}

If we now substitute the parameters
in Eq.~\ref{equ:tl4} by the values we use in this
paper, and using BER$=10^{-7}, 10^{-6}, 10^{-5}$ we
get that $X_{opt}= 0.0991 \cdot R, 0.3117 \cdot R, 0.9678 \cdot R$
respectively. $X_{opt}$ does not depend on the MSDU size
but it is a function of the PHY rate $R$. If we look for
the PHY rates for which $X_{opt} > 64$, i.e. 11ax/256 outperforms
11ax/64, we get the following PHY rates $ 645, 205, 66 Mbps$ 
respectively. This means that the corresponding $MCS_C$s are
{\it MCS2, MCS0, MCS0} respectively, as is shown
in Figures~\ref{fig:comptwole5}-\ref{fig:comptwole7} respectively.
Notice that by the above in turns out that the $MCS_C$s
do not depend on the MSDUs' sizes, as it is also observed from
Figures~\ref{fig:comptwole5}-\ref{fig:comptwole7}.

\section{Summary}

A comparison between the maximum Throughputs of
IEEE 802.11ax and IEEE 802.11ac in a single user
operation mode is performed, in a scenario where
one user transmits continuously to another user using
Two-Level aggregation. Concerning IEEE 802.11ax two
flavors are considered, using acknowledgment
windows of 256 and 64 MPDUs respectively.

IEEE 802.11ax outperforms IEEE 802.11ac by 48$\%$ and 29$\%$
in unreliable and reliable channels respectively.
Also, a detailed analysis comparing between the
two flavors of IEEE 802.11ax is given.

This paper is one of the first to evaluate the performance
of IEEE 802.11ax and more are expected to come for other
scenarios such as the multi user operation mode.

\clearpage

%%%%%%%%%%%%%%%%%%%%%%%%%%%%%%%%%%%%%%%%%%%%%%%%%%%%%%%%%%

\bibliographystyle{abbrv}
\bibliography{main}

%%%%%%%%%%%%%%%%%%%%%%%%%%%%%%%%%%%%%%%%%%%%%%%%%%%%%%%%%%%%%%%%

\end{document}